\documentclass[aps, prl,superscriptaddress,twocolumn,floatfix]{revtex4}
\usepackage{amsmath}
\usepackage{amssymb}
\usepackage{amsfonts}
\usepackage{graphicx}
\usepackage{hyperref}

\begin{document}
\title{$g$-Tensor Control in Bent Carbon Nanotube Quantum Dots}
\author{R.~A.~Lai}
\thanks{Present address: Department of Physics, Stanford University, Stanford, California 94305, USA.}
\affiliation{Department of Physics, Harvard University, Cambridge, Massachusetts 02138, USA}

\author{H.~O.~H.~Churchill}
\thanks{Present address: Department of Physics, Massachusetts Institute of Technology, Cambridge, Massachusetts 02138, USA.}
\affiliation{Department of Physics, Harvard University, Cambridge, Massachusetts 02138, USA}

\author{C.~M.~Marcus}
\email{marcus@nbi.dk}
\affiliation{Department of Physics, Harvard University, Cambridge, Massachusetts 02138, USA}
\affiliation{Center for Quantum Devices, Niels Bohr Institute, University of Copenhagen, 2100 Copenhagen \O, Denmark}

\newcommand{\vsg}{V_{\rm{SG}}}
\newcommand{\vbg}{V_{\rm{BG}}}
\newcommand{\vsd}{V_{\rm{SD}}}
\newcommand{\vs}{V^*}
\newcommand{\pdot}{\varphi_{\rm{max}}}
\newcommand{\pb}{\varphi}
\newcommand{\p}{\varphi}
\newcommand{\ptube}{\varphi_{\rm{tube}}}
\newcommand{\gt}{g}
\newcommand{\pave}{\varphi_{\rm{ave}}}
\newcommand{\pmax}{\varphi_{\rm{max}}}
\newcommand{\bx}{B_{\rm{x}}}
\newcommand{\by}{B_{\rm{y}}}
\newcommand{\ml}{\Delta \mu_{\rm{L}}}
\newcommand{\mr}{\Delta \mu_{\rm{R}}}

\begin{abstract}
We demonstrate gate-control of the electronic $g$-tensor in single and double quantum dots formed along a bend in a carbon nanotube. From the dependence of the single-dot excitation spectrum on magnetic field magnitude and direction, we extract spin-orbit coupling, valley coupling, spin and orbital magnetic moments. Gate control of the $g$-tensor is measured using the splitting of the Kondo peak in conductance as a sensitive probe of Zeeman energy.  In the double-quantum-dot regime, the magnetic field dependence of the position of cotunneling lines in the two-dimensional charge stability diagram is used to infer the real-space positions of the two dots along the nanotube.
\end{abstract}

\maketitle

Carbon nanotubes have several attractive properties that make them favorable candidates for spin qubits and spintronics applications, with many recent experimental advances particularly in the area of few-electron quantum devices~\cite{mason2004local,grove2008triple,jorgensen2008singlet,buitelaar2008pauli,steele2009tunable,hughc132009}. 
In addition to a low concentration of nuclear spins and large confinement energies, nanotubes exhibit a unique circumferential spin-orbit coupling, which has been described theoretically~\cite{izumida2009} and characterized experimentally~\cite{ferdi2008,hugh2009,jespersen2011}. 
This spin-orbit coupling generates an effective magnetic field parallel to the axis of the nanotube.  
Bends in nanotubes couple position and spin by creating a spatial dependence of the direction and magnitude of this effective magnetic field~\cite{flensberg2010}. 
Recently, this effect has been used to facilitate electron dipole spin resonance and qubit manipulation~\cite{pei2012valley,laird2012}.
Here, we demonstrate another key feature of nanotubes with bends: control of the $g$-tensor via electrostatic gates~\cite{flensberg2010}. In addition, we demonstrate the use of $g$-tensor anisotropy to extract the positions of gate-defined single and double quantum dots along a curved  nanotube, taking advantage of the $g$-to-position mapping made possible by a bend.
\vspace{-10pt}

The device was based on a single-walled carbon nanotube grown by methane chemical vapor deposition. During growth, van der Waals forces bind large numbers of nanotubes to the silicon substrate in random orientations, including bent configurations. Though not used here, deliberate bending of nanotubes has also been demonstrated~\cite{falvo1997nanotubes,poncharal1999electrostatic, biercuk2004, geblinger2008self}.
A bent tube was identified by scanning electron microscopy, and Ti/Pd contacts (5/50 nm thick) and a local side-gate (SG) proximal to the left arm of the bend were patterned by electron beam lithography [Fig.~1(a)]. 
The degenerately doped silicon substrate formed a global back-gate (BG) insulated by $0.5\,\mu$m of thermal oxide.
The radius of curvature was $\sim\,0.5\, \mu$m, with the two arms of the bend at angles $130^{\circ}$ and $0^{\circ}$ relative to the x-axis. 
The device was measured in a dilution refrigerator with an electron temperature of 0.1~K using direct current and lock-in techniques with a vector magnetic field $\vec{B}$, which we label in polar and cartesian coordinates in the x-y plane of the tube as $\vec{B}=(B,\p)=(\bx,\by)$.

\par
\begin{figure}[b!]
\center \label{figure1}
\includegraphics[width=3.4in]{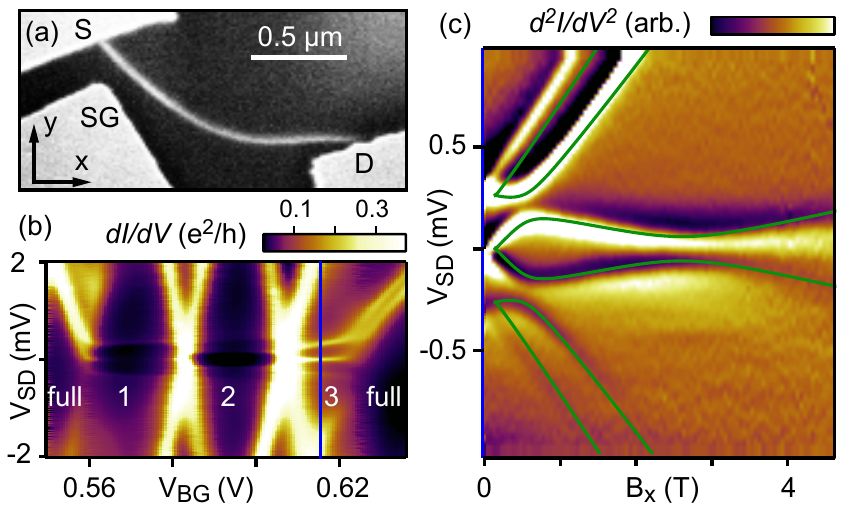}
\caption{\footnotesize{(a) SEM micrograph of the bent carbon nanotube device, consisting of Pd source (S) and drain (D) contacts, a local side-gate (SG), and a global back-gate (BG).   (b) Differential conductance $dI/dV$ as a function of the back-gate voltage $\vbg$ and source-drain bias, $\vsd$, at $B=0$ T.
(c) Cotunneling spectroscopy in $d^2I/dV^2$ as a function of a magnetic field $\bx$ in the plane of the chip at $\vbg=0.615$V, corresponding to the blue line in (b).  Green curves are fits to the model of the electronic spectrum of straight nanotubes including spin (43 $\mu$eV/T) and valley (290 $\mu$eV/T) magnetic moments, spin-orbit (220 $\mu$eV) and valley (70 $\mu$eV) couplings, magnetic field relative to the nanotube (15$^{\circ}$), as described in Ref.~\cite{jespersen2011}, and a threshold magnetic field ($0.15$~T), as described in Refs.~\cite{kogan2004measurements, costi2000kondo}.}}
\end{figure}

Coulomb diamonds with fourfold shell structure were observed in differential conductivity, $dI/dV$, as a function of source-drain bias, $\vsd$, and back-gate voltage, $\vbg$, with the side-gate grounded, as shown in Fig.~1(b)~\cite{jarillo2005orbital,liang2002}.
A group of four resonances closely spaced in $\vbg$, corresponding to consecutive filling of the spin-valley levels of a longitudinal orbital shell, can be seen in Fig.~1(b). A repeating pattern of three-two-three conductance ridges for $\vsd <$ 0.5 mV are seen at occupancies of one, two, and three electrons, respectively, above a full shell, which we attribute to the Kondo features in even and odd occupied Coulomb blockade valleys. 

The single-dot spectrum as a function of $\bx$ was extracted from conductance data in the Kondo regime at $\vbg=0.616$~V and $\by = 0$, using the inflection point of conductance as a function of $\vsd$ to measure level position~\cite{rosch2003nonequilibrium}, as shown in Fig~1(c).
The dependence of the spectrum on $\bx$ can be understood as resulting from a combination of spin-orbit coupling and/or valley mixing that breaks the degeneracy of the four spin-valley states, consistent with theory~\cite{fang2008} and previous experiment~\cite{jespersen2011}. 
Fitting the spectrum at negative bias to a model of a locally straight nanotube~\cite{jespersen2011} yields spin (43 $\mu$eV/T) and valley (290 $\mu$eV/T) magnetic moments, spin-orbit (220 $\mu$eV) and valley (70 $\mu$eV) couplings, and magnetic field angle relative to the nanotube (15$^{\circ}$) [Fig.~1(c)]. The model also includes a threshold magnetic field of $0.15$~T of Kondo peak splitting~\cite{kogan2004measurements, costi2000kondo}.  We observed that the cotunneling spectrum is asymmetric in bias, as seen previously~\cite{Amasha2005Kondo}. 
Device parameters are consistent with previous measurements~\cite{ferdi2008, hugh2009, jespersen2011}, except for the spin $g$-factor (1.5), which is somewhat smaller than previously 
reported ($g=2$).

\begin{figure}
\center \label{figure2}
\includegraphics[width=3.4in]{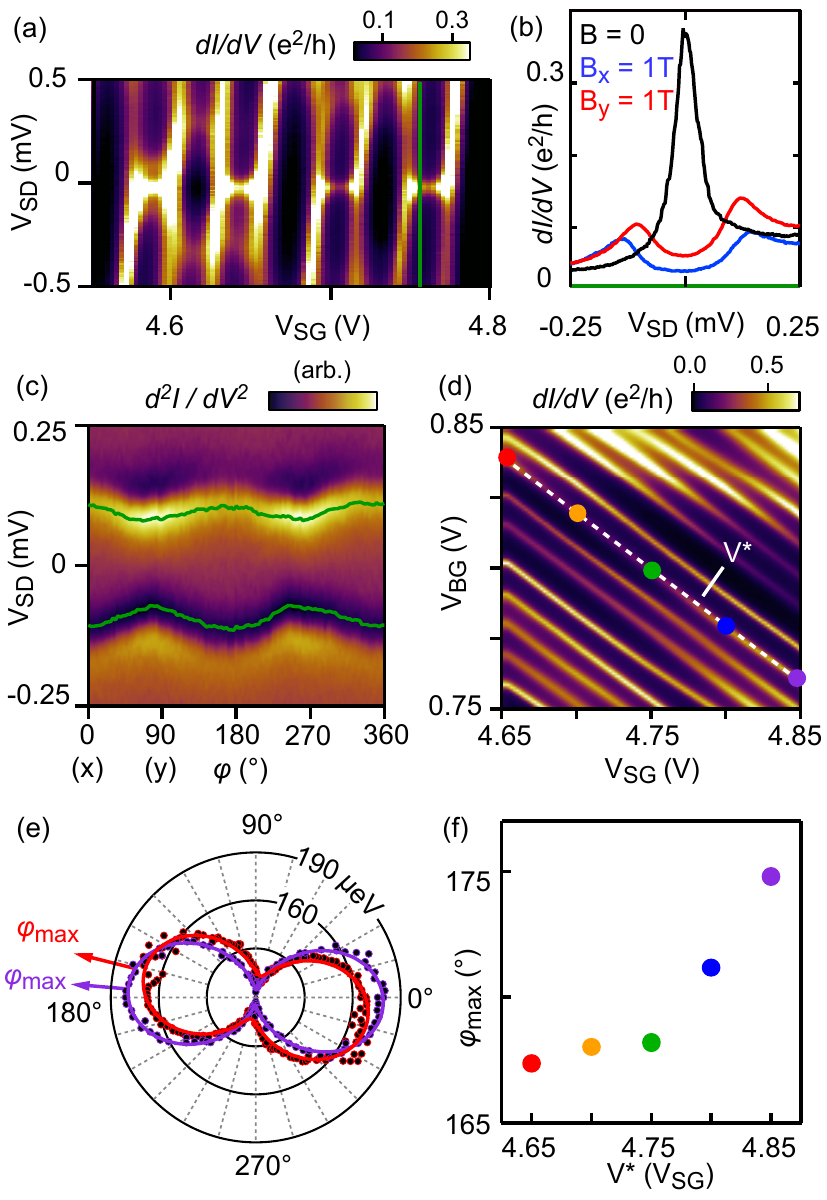}
\caption{\footnotesize{
(a)  $dI/dV$ with $\vbg=0.8V$ as a function of bias and $\vsg$ showing Coulomb diamonds in the Kondo regime.  
(b) Bias dependence of $dI/dV$ with $\vsg=4.76$ V (indicated by the green line, (a)) at $B=0$ (black trace), $\bx=1$ T (blue), and $\by=1$ T (red).  
(c) $d^2I/dV^2$ at the gate voltages indicated by the green dot in (a) as a 1 T magnetic field is rotated in the x-y plane.  The green line tracks the inflection point position indicated by the first maximum and minimum at positive and negative bias, respectively.  
(d) $dI/dV$ as a function of $\vsg$ and $\vbg$ at zero magnetic field.   We define the voltage $\vs$, parametrized by $\vsg$, along a single charge transition shown by the dashed white line.   
(e) The inflection point separation (dots), sinusoidal fit to the splitting (solid curves), and angle of maximum Kondo peak splitting $\pmax$ (arrows) are shown in red and purple at values of $\vs$ corresponding to the red and purple dots in (a), respectively. 
(f) $\pmax$ extracted from sinusoidal fits as in (e) for the five settings of $\vs$ shown in (a).}}
\end{figure}

Differential conductance, $dI/dV$, as a function of $\vsd$ and $\vsg$ with $\vbg=0.72$ V showed Coulomb diamonds with Kondo ridges at zero bias for every other charge state [Fig.~2(a)], similar to those in Fig.~1(b). 
In this regime, the Kondo peak shows a larger splitting for $\bx$ than for $\by$ [Fig.~2(b)], indicating that $g$ is anisotropic. Anisotropy of the $g$-tensor can also be seen in a plot of $d^2I/dV^2$ as function of bias and magnetic field angle in the plane [Fig.~2(c)]. Here, the two inflection points of conductance appear as maxima and minima at positive and negative bias, respectively, with a maximal splitting near  $\pb \sim 0^{\circ} (\rm{x})$ and a minimal splitting near $\pb \sim 90^{\circ} (\rm{y})$.

Control of the $g$-tensor is achieved by moving the position of the dot along the nanotube bend without changing its occupancy using two gates acting in opposition. For a many-electron quantum dot ($N\sim$ 70), we can introduce a single voltage axis, $\vs$, parametrized by $\vsg$, that tracks a Coulomb resonance as both gate voltages (SG and BG) are swept, as shown in Fig.~2(d).  
At all values of $V^{*}$, we observed a sinusoidal dependence of the Kondo splitting on magnetic field angle in the plane [Fig.~2(c)], with the amplitude and phase of the sinusoid depending on $V^{*}$. As an example, Kondo splittings along with sinusoidal fits are shown in a polar plot in Fig.~2(e) for  $\vs$=4.65~V (red), corresponding to $\vsg=4.65$~V and $\vbg=0.84$~V, and $\vs$=4.85~V (purple), corresponding to $\vsg=4.85$~V and $\vbg=0.76$~V. The angles of maximal splitting, $\pmax$, are clearly different for the two cases. Figure~2(f) shows values for $\pmax$ for five values of $\vs$, corresponding to the points marked in Fig.~2(a). The angle where the maximum splitting occurred was found to change monotonically with a rough rate of $\sim8^{\circ}$ for a change in $\vs$ of 0.2 V.
Based on the increase in $\pmax$ for increasing $\vs$, we note that in this instance the dot moved toward the right side of the bend as $\vs$ became more positive (while the global back gate became more negative). Given the position of the side gate, one might have expected the opposite direction of motion. Due to disorder, however, the dot can move either way as a function of $\vs$. Fabricating multigate devices and reducing disorder will both serve to increase control of dot position. 

Additional features associated with a bend are evident when the device is tuned to form a double quantum dot. 
This regime is characterized by the familiar honeycomb charge stability diagram, as seen in Fig.~3(a). 
To allow energy level shifts as a function of magnetic field to be examined for the two dots separately, we define two axes, $V1$ and $V2$, [dotted lines in Fig.~3(a)] midway between adjacent triple-point resonances.
Changes in the position of points L and R, where axes $V1$ and $V2$ cross the cotunneling lines that define the charge stability boundaries, were used to track the evolution of energy levels of the left and right dots. (The left dot is the one closest to the side gate.)
Accurate positions for points L and R were found by fitting cotunneling peaks along $V1$ and $V2$ to the form $I(V)=I_0\cosh^{-2}\left[(V-V_0)/2W\right]$ [Fig.~3(a)~inset], where $W$ is the peak width due to both temperature and tunneling, expected to be valid for $\vsd \ll W$~\cite{beenakker1991}.
The observed behavior is well described by a capacitance model of a double quantum dot~\cite{vanderwiel} and gives the tunnel coupling $t=0.59$~meV and the mutual charging energy $U_{\rm{Cm}}=1.2$~meV. 
From the size of the bias triangles at the triple-point between the points L and R in finite-bias measurements, we obtain the gate lever-arms that convert gate voltage and energy~\cite{alphas}.
We note that the same lever arm values were found at $\bx=0.5$~T and $1.5$~T.

\begin{figure}
\center \label{figure3}
\includegraphics[width=3.3in]{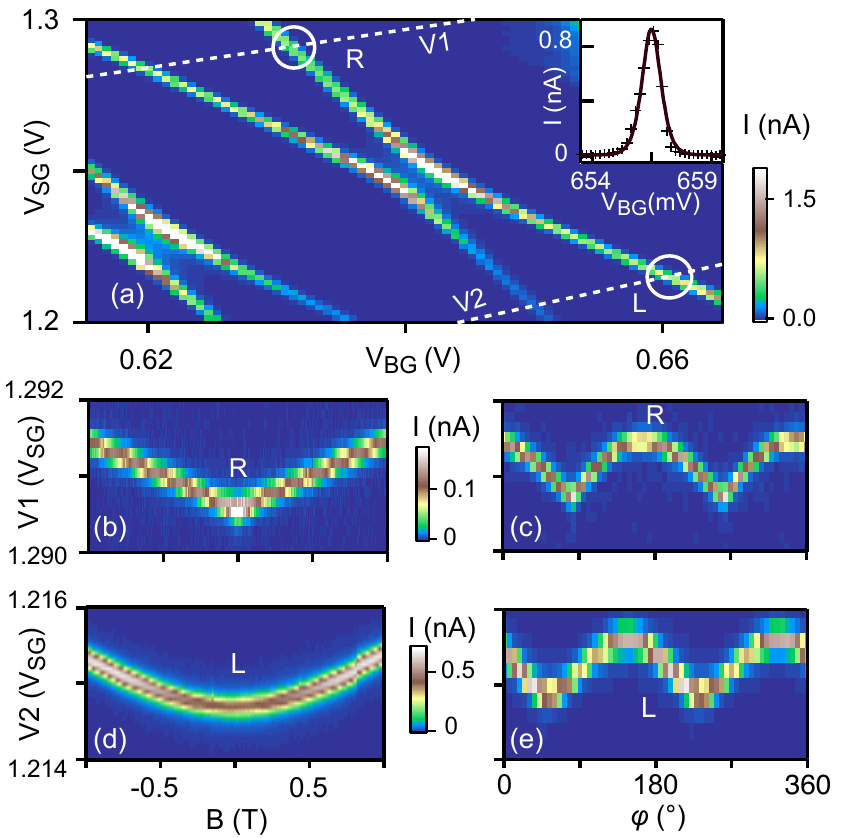}
\caption{\footnotesize{(a) Current as a function of $\vsg$ and $\vbg$ showing the honeycomb charging pattern of a few-electron double-dot at $B=0$ and $\vsd=0$. Voltages $V1$ and $V2$ (parametrized by $\vsg$) cut across the cotunneling transitions L, R of the double dot, maximally detuned from the triple-point resonances. Inset: Crosses are the current as a function of $\vbg$ at $\vsg=1.223$~V. The solid line is the fit to a $\cosh^{-2}$ lineshape (see text). 
(b), (c) Current along $V1$ as a function of magnetic field magnitude at $\pb=0$ (b) and in-plane angle at $B=1$~T (c). (d), (e) Same as (c), (d) along $V2$.}}
\end{figure}

Cotunneling current along $V1$ and $V2$ as a  function of $B$ and $\pb$ are shown in Figs.~3(b-e).
The dependence of points L and R on $\pb$ reflects anisotropy in $\gt$ for the left and right dots, respectively. In particular, larger shifts from the zero field values occur when the field is parallel to the section of the nanotube containing the dot; smaller shifts occur when the field is  perpendicular to the nanotube segment. We note that the positions of the extrema in this dependence for the left and right dots do not occur at the same angles. From these dependences, we conclude that the left and right dots reside in segments of the tube that make different angles with respect to the applied field. That is, the double dot is on a bend. 

Two-dimensional plots showing the movement of points R and L from their zero-field values, defined as $\Delta V1$ and $\Delta V2$, as a function of both field angle and magnitude are shown in Figs.~4(a,b).
Field direction dependences of $\Delta V1$ and $\Delta V2$ follow similar evolution, with an offset in phase. We also observe that there does not exist a field angle about which the level shifts are symmetric, especially evident for $B<0.5$~T. 
We interpret the lack of symmetry, which gives Figs.~4(a,b) their overall canted appearance, as indicating that each dot extends along a segment of bent tube. The bend breaks the symmetry that would be present if each dot were within straight segment of tube. 

To model the angle dependence of $\Delta V1$ and $\Delta V2$, we assume the energy levels of each dot respond to the applied field as a function of the difference of the field angle, $\pb$, and the nanotube angle of a completely localized dot, $\pdot$.
Thus, the energy levels for the right and left dots in gate voltage respond periodically with magnetic field angle, with a maximum corresponding to parallel field when $\pb=\pdot$ and a minimum corresponding to perpendicular field, 
$\Delta V1(2) \propto \left| \cos \frac{\pi}{180}\left(\pb-\pdot^{R(L)}\right) \right|$. 
We expect the model to be valid at higher fields, when the orbital magnetic energy is larger than the spin Zeeman energy, valley scattering, spin-orbit coupling, and changes in the charging energy. 

Figure 4(c) shows good agreement between data at $B=1$~T, fit using the $\cosh^{-2}$ form given above, and the model of phase evolution.
Deviations between data and model may reflect the finite extent of the electron distribution or shifts in that distribution with magnetic field angle.
We interpret $\pdot$ as the tangent angle of the nanotube at the mean position of each quantum dot. With this interpretation, we can convert $\pdot$ to a position in real space using the angle distribution of the nanotube~[Fig.~4(d)~inset] and the micrograph of the device~[Fig.~4(d)].
The positions of the two dots represented by $\pdot$ are shown as a red circle ($\pdot^L=143^{\circ}$) and a purple circle ($\pdot^R=167^{\circ}$) in Fig.~4(d).

We conclude that by using a bend in a carbon nanotube, both the magnitude and angular orientation of the $g$-tensor can be controlled using electrostatic gates. Future experiments could use these effects to produce rapid spin manipulation beyond EDSR, for instance, by initializing a spin in one dot and moving non-adiabatically to the other, resulting in spin rotation at the Larmor frequency. 

\begin{figure}
\center \label{figure4}
\includegraphics[width=3in]{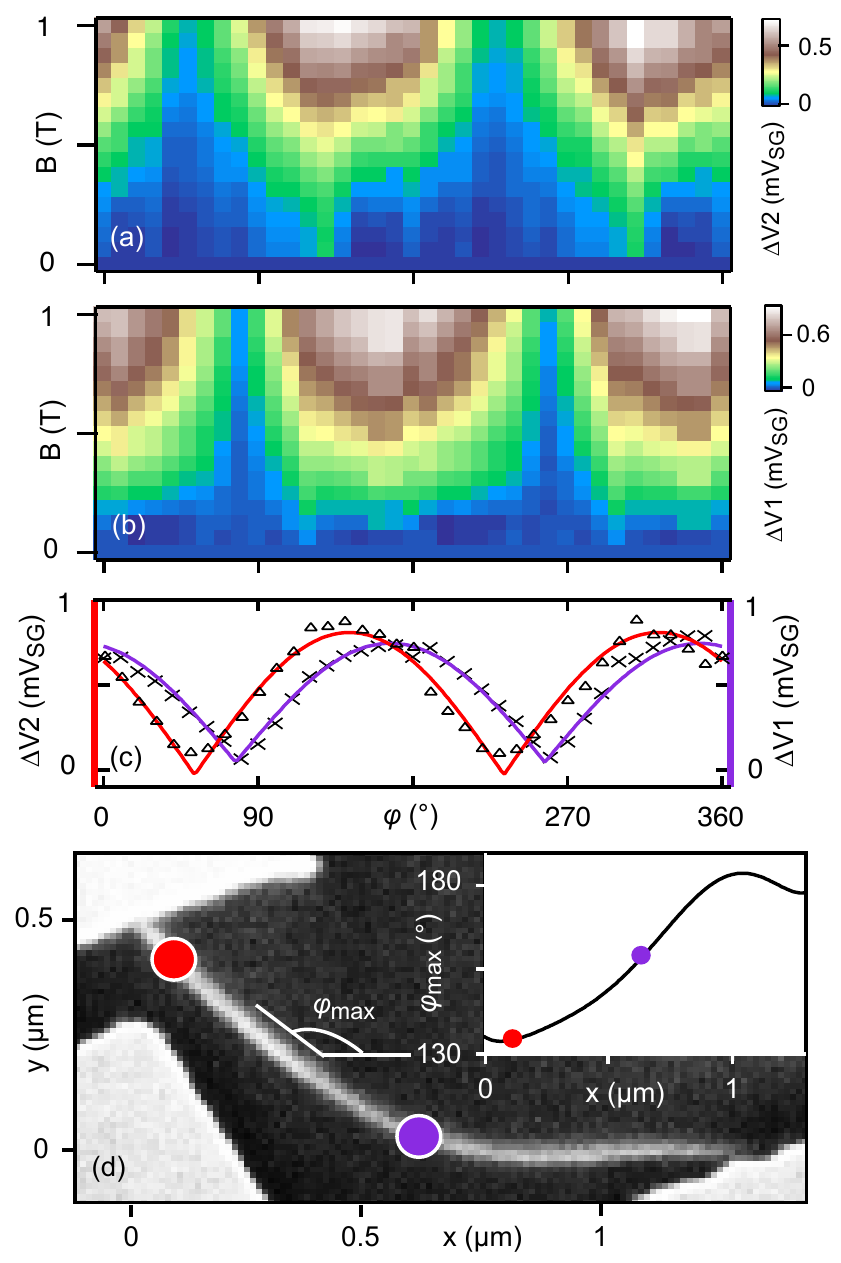}
\caption{\footnotesize{(a)~$\Delta V2$, the shift of the point L along $V2$ due to the vector magnetic field. (b)~The same as (a) for $\Delta V1$, the shift of the point R along $V1$. $\Delta V1$ and $\Delta V2$ are parametrized by $\vsg$. (c)~Black crosses show the shift of L, $\Delta V2$, at $B = 1 T$ (left axis). Black triangles show the shift of R, $\Delta V1$, at $B = 1$~T (right axis). Red and purple traces are the fits to $\left|\cos\frac{\pi}{180^{\circ}}\left(\p-\pdot\right)\right|$ for the points L and R, respectively.
(d)~Inset: the distribution of the angle tangent to the nanotube along the x direction, measured as shown in the white diagram on the SEM micrograph.
Red and purple circles in the inset show $\pdot$ for the left and right dots, respectively. 
Associating $\pdot$ with the measured nanotube tangent angles, we infer the average position of the left and right dots, shown as the red and purple circles, respectively, on the micrograph.}}
\end{figure}

Support from the National Science Foundation through the Materials World Network and NRI, through the INDEX Center is acknowledged. We also thank IBM, Harvard University, and the Danish National Research Foundation for support. 

\bibliography{Refs}

\begin{thebibliography}{27}
\bibitem[{alp()}]{alphas}
\bibinfo{note}{Lever arms for the left and right dots to the side gate (SG) and
  back gate (BG) are given by $\alpha^{L}_{BG}=0.49e, \alpha^{L}_{SG}=0.27e,
  \alpha^{R}_{BG}=0.51e,\alpha^{R}_{SG}=0.13e$}.
\end{thebibliography}

\end{document}